\documentclass[aps,prl,twocolumn,amsmath,amssymb,superscriptaddress,longbibliography]{revtex4-2}
\usepackage{blindtext}
\usepackage{enumitem}
\usepackage{graphicx}
\graphicspath{{figures/}}

\usepackage{dcolumn}
\usepackage{color}
\usepackage{xcolor}
\usepackage{amsmath}
\usepackage{braket}

\usepackage{bm}
\usepackage{changes}
\usepackage{chemformula}

\usepackage[breaklinks,colorlinks,bookmarks=false,citecolor=blue,linkcolor=red,urlcolor=blue]{hyperref}
\usepackage{cleveref}
\crefname{equation}{Eq.}{Eqs.}
\Crefname{equation}{Eq.}{Eqs.}
\crefname{figure}{Fig.}{Figs.}
\Crefname{figure}{Fig.}{Figs.}
\crefname{section}{Sec.}{Secs.}
\Crefname{section}{Sec.}{Secs.}

\usepackage{amsmath}

\begin{document}
\title{Fragmented Cooper pair condensation in striped superconductors}
\author{Alexander Wietek}
\email{awietek@pks.mpg.de}
\affiliation{Center for Computational Quantum Physics, Flatiron Institute, 162 Fifth Avenue, New York, NY 10010, USA}
\affiliation{Max Planck Institute for the Physics of Complex Systems, N\"othnitzer Strasse 38, Dresden 01187, Germany}
\date{\today}
\begin{abstract}
Condensation of bosons in Bose-Einstein condensates or Cooper pairs in superconductors refers to a macroscopic occupation of a few single- or two-particle states. A condensate is called "fragmented" if not a single, but multiple states are macroscopically occupied. While fragmentation is known to occur in particular Bose-Einstein condensates, we propose that fragmentation naturally takes place in striped superconductors. To this end, we investigate the nature of the superconducting ground state realized in the two-dimensional $t$-$t^\prime$-$J$ model. In the presence of charge density modulations, the condensate is shown to be fragmented and composed of partial condensates located on the stripes. The fragments of the condensates hybridize to form an extended macroscopic wave function across the system. The results are obtained from evaluating the singlet-pairing two-particle density matrix of the ground state on  finite cylinders computed via the density matrix renormalization group (DMRG) method. Our results shed light on the intricate relation between stripe order and superconductivity in systems of strongly correlated electrons.  
\end{abstract}
\maketitle

\section{Introduction}
\label{sec:introduction}
Superconductivity constitutes one of the most fascinating ramifications of
quantum mechanics in macroscopic condensed matter systems. 
A key role in our understanding of high-temperature superconductivity is attributed
to the two-dimensional Hubbard model, or its strong coupling limit, the $t$-$J$
model~\cite{Anderson1987,Zhang1988,Emery1988}. Early on it was realized, that the
essential behavior of the copper-oxide
superconductors might be captured by these basic models. Solving these models,
however, has posed major difficulties which have fueled the
development of sophisticated numerical and analytical methods over the last
decades~\cite{Imada1998,Qin2021,Arovas2022}. 
These efforts have led to considerable progress in recent
years~\cite{tremblay2006review,maier2005review,Qin2021,Simkovic2021,Schaefer2021}. The emergence
of stripes in certain relevant regions of the phase diagram, first proposed by
Hartree-Fock studies~\cite{Zaanen1989,Poilblanc1989,Machida1989,Kato1990}, has by now been firmly
established by a broad range of numerical
methods~\cite{Martins2000,LeBlanc2015,Zheng2017,Huang2017,Huang2018,Wietek2021}. 
The more intricate question of whether superconductivity is realized at low 
temperature in these models is currently being tackled by various approaches~\cite{Qin2020}.
This year, three density matrix renormalization group
(DMRG)~\cite{White1992,White1993} studies have reported 
robust d-wave superconductivity in particular regimes of the hole-doped
$t$-$t^\prime$-$J$ model~\cite{Gong2021,Jiang2021,HCJiang2021}. 
Refs.~\cite{Gong2021,HCJiang2021} employed advanced large-scale DMRG simulations
to achieve convergence towards power-law decay of
superconducting pairing correlations, indicative of a quasi-1D descendant of a 2D
superconductor. Ref.~\cite{Jiang2021} applied pinning fields, to demonstrate strong d-wave pairing in an extended region of the phase diagram. 

In this manuscript, we investigate the nature of the superconducting condensate in this model in
further detail. We propose to study the eigenvalues and eigenvectors of
a properly chosen two-particle density matrix, which describe the superconducting
condensate fraction and the macroscopic condensate wave function. The method is 
applied to the superconducting ground state of the $t$-$t^\prime$-$J$ model 
obtained from DMRG on cylinders of width $W=4$ and $W=6$. We discover, that in 
the presence of stripes, not just one global superconducting condensate but multiple condensates are formed. Each partial condensate is found to be associated with a single charge stripe. The occurrence of multiple condensates,
corresponding to multiple dominant eigenvalues of the two-body density matrix is
called \textit{fragmentation}. Fragmentation is known to occur in specific instances of Bose-Einstein condensates~\cite{Spekkens1999,Ho2000,Mueller2006,Kang2014}. Examples include in weakly-interacting spinor condensates~\cite{Evrard2021} and exciton condensates~\cite{Combescot2015}. However, fragmentation has to the best of our knowledge not prominently been discussed in the context of high-temperature superconductivity. The method of studying eigenvalues and eigenvectors of a two-particle density matrix is applicable for any numerical method, but particularly well-suited for DMRG. We, therefore, suggest this approach as a reliable means of diagnosing superconductivity in correlated electron systems. 

\section{Two-particle density matrices}
\label{sec:condensate}

The essential quantity to study condensation of Cooper pairs is the generic
two-particle density matrix $\rho_2$~\cite{Legett2006}, 
\begin{equation}
\label{eq:generaltwobodydm}
    \rho_2(\bm{r}_i\sigma_i, \bm{r}_j\sigma_j|
    \bm{r}_k\sigma_k,\bm{r}_l\sigma_l) = \langle
    c^\dagger_{\bm{r}_i\sigma_i} c^\dagger_{\bm{r}_j\sigma_j}
    c_{\bm{r}_k\sigma_k} c_{\bm{r}_l\sigma_l}
    \rangle,
\end{equation}
where $\sigma_i = \uparrow, \downarrow$ denotes the fermion spin and $c^\dagger_{\bm{r}_i\sigma_i}$ and $c_{\bm{r}_i\sigma_i}$ are fermion creation and annihilation operators at lattice positions $\bm{r}_i$.
Since $\rho_2$ is Hermitian, 
\begin{equation} 
    \label{eq:hermitecity}
    \rho_2(\bm{r}_i\sigma_i, \bm{r}_j\sigma_j|
    \bm{r}_k\sigma_k,\bm{r}_l\sigma_l)  = 
    \rho_2^*(\bm{r}_k\sigma_k, \bm{r}_l\sigma_l|
    \bm{r}_i\sigma_i,\bm{r}_j\sigma_j),
\end{equation}
it can be diagonalized with real eigenvalues $\varepsilon_n$ and eigenvectors $\chi_n$,
\begin{align}
    \label{eq:dmatdiagonal}
    \begin{split}
        \rho_2(\bm{r}_i\sigma_i, \bm{r}_j\sigma_j|
    \bm{r}_k\sigma_k,\bm{r}_l\sigma_l) =& \\
    \sum_{n} \varepsilon_n 
    \chi_n^* (\bm{r}_i\sigma_i, \bm{r}_j\sigma_j)&
    \chi_n (\bm{r}_k\sigma_k, \bm{r}_l\sigma_l).
    \end{split}
\end{align}
In analogy to Bose-Einstein condensation, Cooper pair condensation takes place 
whenever one or more eigenvalues are of order $N$, where $N$ is the number of lattice
sites. If exactly one eigenvalue is of order $N$ the condensate is referred 
to as \textit{simple}. If more than one eigenvalue is of order $N$, the condensate
is called \textit{fragmented}~\cite{Legett2006}. Dominant eigenvalues $\varepsilon_i$ are referred to as the condensate fractions. 

\begin{figure}[t]
    \centering
    \includegraphics[width=\columnwidth]{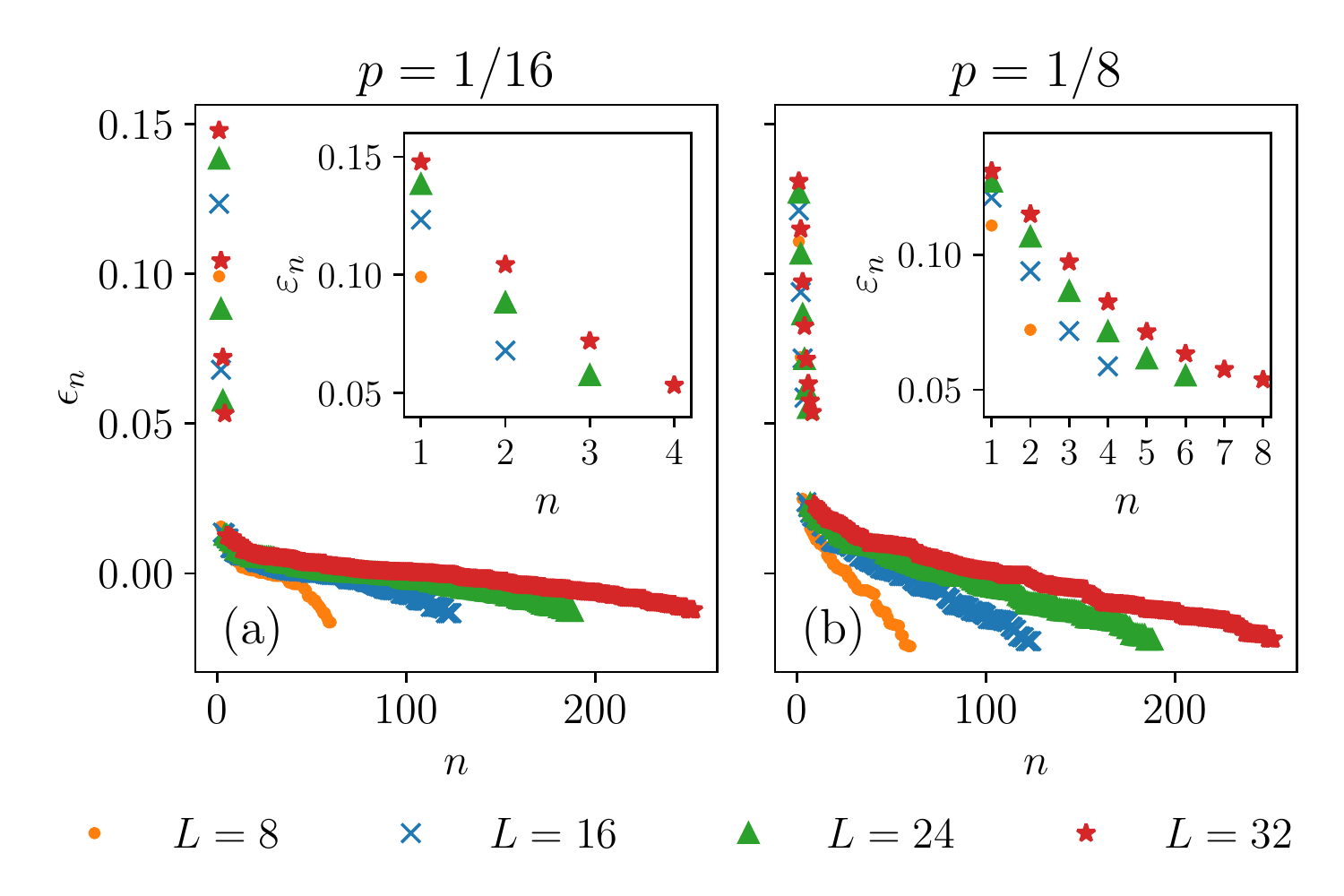}
    \caption{Spectrum $\varepsilon_n$ of the singlet density matrix $\hat{\rho}_S$ of the ground state on the width $W=4$ cylinder at $t^\prime=0.2$ and $J=0.4$. We compare system lengths $L=8, 16, 24, 32$ and show results for hole-doping $p=1/16$ (a) and $p=1/8$ (b). The number of dominant eigenvalues above the residual continuum exactly matches the number of stripes in the system. The insets zoom in on the largest eigenvalues. The condensate fractions $\varepsilon_n$ increase with system size.}
    \label{fig:pairing_cond_eigs}
\end{figure}

\begin{figure}[t]
    \centering
    \includegraphics[width=\columnwidth]{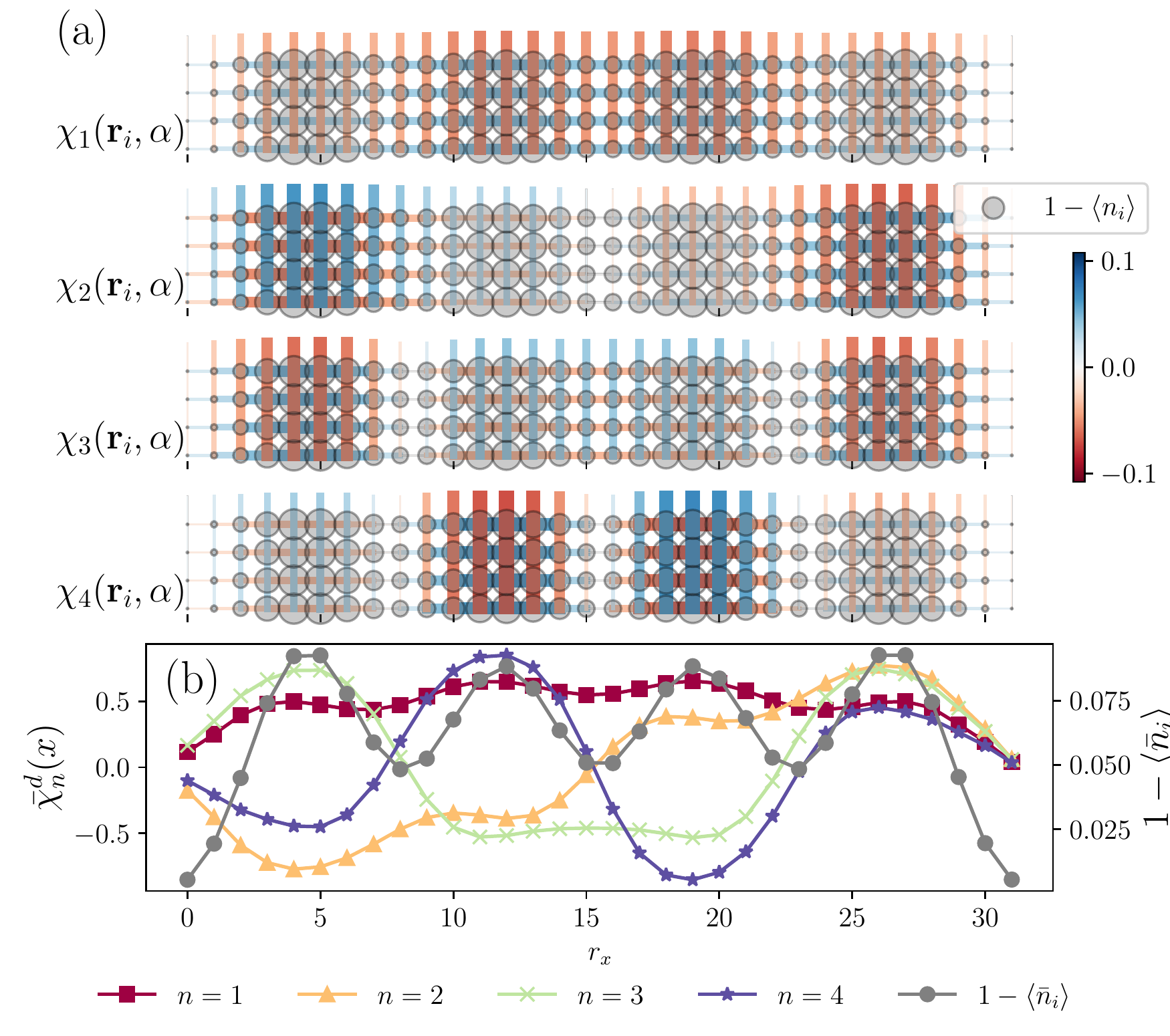}
    \caption{(a) Condensate wave functions $\chi_n(\bm{r}_i, \alpha)$ for the dominant
    four eigenvalues of the two-body density matrix on a $32 \times 4$ cylinder at
    doping $p=1/16$ and $t^\prime= 0.2$. For $\alpha = \hat{\bm{x}}$ we show 
    the value of $\chi_n(\bm{r}_i, \alpha)$ as the color and line width right
    to the site $\bm{r}_i$, for $\alpha = \hat{\bm{y}}$ it is shown on the link
    on top of site $\bm{r}_i$. Blue (red) indicates a positive (negative) value of
    $\chi_n(\bm{r}_i, \alpha)$. The hole-density $1 - \langle n_i \rangle$ is shown as
    the area of the gray circles. We observe uniform d-wave pattern, where vertical and horizontal bonds have opposite signs in the most dominant condensate wave function, while the other dominant condensates exhibit modulation of the d-wave orientation concomitant with the stripes. (b) Rung-averaged $d$-wave condensate wave function $\bar{\chi}_i^d$ and the rung-averaged hole-density $1-\langle\bar{n}_i\rangle$.}
    \label{fig:w4condwfs}
\end{figure}

While the above definitions are rather generic in scope, we focus on more
specific quantities to investigate singlet-pairing in two-dimensional lattice models. First, we define the singlet-pairing density matrix $\rho_S$ as, 
\begin{equation}
\label{eq:singletpairingmatrix}
\rho_S(\bm{r}_i, \bm{r}_j | \bm{r}_k ,\bm{r}_l) = 
\langle \Delta_{\bm{r}_i\bm{r}_j}^\dagger \Delta_{\bm{r}_k\bm{r}_l}\rangle,
\end{equation}
where the singlet-pairing operators $\Delta_{\bm{r}_i\bm{r}_j}$ is given by
\begin{equation}
\label{eq:singletpairingop}
\Delta_{\bm{r}_i\bm{r}_j}^\dagger= \frac{1}{\sqrt{2}}
\left( 
c^\dagger_{\bm{r}_i\uparrow} c^\dagger_{\bm{r}_j\downarrow} - 
c^\dagger_{\bm{r}_i\downarrow} c^\dagger_{\bm{r}_j\uparrow}
\right).
\end{equation}
To focus on two-dimensional lattice geometries, we consider a nearest-neighbor singlet density matrix,
\begin{equation}
\label{eq:nnsingletpairingmatrix}
\rho_S(\bm{r}_i, \alpha | \bm{r}_j, \beta) = \rho_S(\bm{r}_i ,(\bm{r}_i + \alpha) | \bm{r}_j, (\bm{r}_j + \beta)),
\end{equation}
where $\alpha$ (resp. $\beta$) denote the vectors connecting nearest-neighbors on the
lattice, e.g. $\alpha = \hat{\bm{x}},  \hat{\bm{y}}$ in the case of a square lattice.
Again, this matrix can be decomposed into eigenvectors, 
\begin{equation}
\label{eq:nnsingletpairingmatrixdecomp}
\rho_S(\bm{r}_i, \alpha | \bm{r}_j, \beta) = 
\sum_n \varepsilon_n \chi_n^*(\bm{r}_i, \alpha) \chi_n(\bm{r}_j, \beta).
\end{equation}
The eigenvectors $\chi_n(\bm{r}_i, \alpha)$ are also referred to
as \textit{macroscopic wave functions}. They depend only on the position $\bm{r}_i$ and the
direction of the nearest-neighbor $\alpha$. In order to exclude local contributions from density and spin correlations, we consider the non-local singlet density matrix,
\begin{align}
\label{eq:nonlocalmatrix}
\begin{split}
    \hat{\rho}_S&(\bm{r}_i, \alpha | \bm{r}_j, \beta) =  \\
&=\begin{cases}
\rho_S(\bm{r}_i, \alpha | \bm{r}_j, \beta) &\mbox{if } 
\{ \bm{r}_i, \bm{r}_i + \alpha \} \cap \{ \bm{r}_j, \bm{r}_j + \beta \} = \emptyset \\
0 &\mbox{else.}
\end{cases}
\end{split}
\end{align}
We note, that with this choice, $\hat{\rho}_S$ is not necessarily positive definite. Thus, eigenvalues of $\hat{\rho}_S$ can in general be positive or negative.

\section{Superconductivity in the $t$-$t^\prime$-$J$ model}
\label{sec:ttpj}

We now investigate the properties of the condensate fractions $\varepsilon_n$ and macroscopic wave functions $\chi_n(\bm{r}_i, \alpha)$ of (non-)superconducting stripe states emerging in a simple model system of strongly interacting electrons. To this end, we study the two-dimensional $t$-$t^\prime$-$J$ model,
\begin{align}
\label{eq:tjmodel}
\begin{split}
 H = &-t \sum_{\langle ij \rangle, \sigma}
    c_{i\sigma}^\dagger c_{j\sigma} + \text{H.c.}
    -t^\prime\sum_{\langle\langle ij \rangle\rangle, \sigma}
     c_{i\sigma}^\dagger c_{j\sigma}  + \text{H.c.}  \\
    &+ J \sum_{\langle ij \rangle}
    \left( \vec{S}_i\cdot \vec{S}_j - \frac{1}{4}n_i n_j \right),
\end{split}
\end{align}
on a square lattice ($c_{i\sigma}^\dagger, c_{i\sigma} = c^\dagger_{\bm{r}_i\sigma}, c_{\bm{r}_i\sigma}$). Here, $\vec{S}_i = (S^x_i, S^y_i, S^z_i)$ are the spin operators, and $n_i = \sum_\sigma c_{i\sigma}^\dagger c_{i\sigma}$ denotes the local density operator. The sums over $\langle i, j \rangle$ are over nearest-neighbor sites
and $\langle\langle i, j \rangle\rangle$ denotes a sum over 
next-nearest neighbors. The Hilbert space is constrained
to prohibit doubly occupied configurations. In the following, we set $t=1$ and 
$J=0.4$ which is the same set of parameters chosen in Ref.~\cite{Jiang2021}. Our
model slightly differs from the model studied in Ref.~\cite{Gong2021}, where also
next-nearest neighbor Heisenberg interactions have been included. While superconductivity of the ground states in particular parameter regimes has already been established~\cite{Gong2021,Jiang2020,Jiang2021,HCJiang2021}, a detailed investigation of two-body density matrices has not previously been performed. 

We apply the DMRG method to study the system on cylindrical geometries 
with open boundary conditions along the long $x$-direction and periodic boundary
conditions along the short $y$-direction. The length in the $x$-direction is denoted by $L$, and the width in the $y$-direction by $W$.
Previous DMRG studies of \cref{eq:tjmodel} have achieved ground state simulations of widths of $W=8$~\cite{Jiang2021}. In this manuscript, we focus on the particular cases of $W=4,6$, which do not require large computational resources to achieve convergence for the ground state. Thus, our computations are less challenging as $W=8$ and, therefore, more easily reproducible. The results in this manuscript have been attained with bond dimensions up to $D=2000$.

We first focus on the case of width $W=4$ cylinders and choose $t^\prime=0.2$ with
hole-dopings $p=1/16$ and $p=1/8$. A previous DMRG study of this model on the width $W=4$ has established an approximate phase
diagram~\cite{Jiang2020}.
For  $t^\prime=0.2$ with hole-dopings
$p=1/16$ and $p=1/8$ the system has been found to exhibit a Luther-Emery liquid 
(the LE2 phase in Ref.~\cite{Jiang2020}) in this regime, with half-filled charge
stripes and pronounced algebraic superconducting correlations.

We computed the singlet density matrix $\hat{\rho}_S(\bm{r}_i, \alpha | \bm{r}_j, \beta)$ by measuring the respective pairing correlations of the ground state
obtained via DMRG. The eigenvalues of the singlet density matrix are for 
cylinder lengths $L=8,16,24,32$ in \cref{fig:pairing_cond_eigs}. The key observation
is that only few dominant eigenvalues are separated from a continuum of minor 
eigenvalues. At hole-doping $p=1/16$ shown in (a), the system realizes one stripe 
for $L=8$, two stripes for $L=16$, three stripes for $L=24$, and four stripes for $L=32$, as can be seen for
$L=32$ in \cref{fig:w4condwfs}. Correspondingly, we observe exactly one dominant
eigenvalue for $L=8$, two dominant eigenvalues for $L=16$, three dominant
eigenvalues for $L=24$, and four dominant eigenvalues for $L=32$. Hence, the number
of dominant eigenvalues exactly matches the number of stripes in the system. The 
same observation is made at hole-doping $p=1/8$,
where twice as many stripes are observed alongside twice as many dominant
eigenvalues. These eigenvalues are interpreted as superconducting condensate
fractions. The insets show a zoom on the dominant eigenvalues $\varepsilon_n$. 
In all cases, the condensate fraction increases monotonously with system size.

\begin{figure}[t]
    \centering
    \includegraphics[width=0.9\columnwidth]{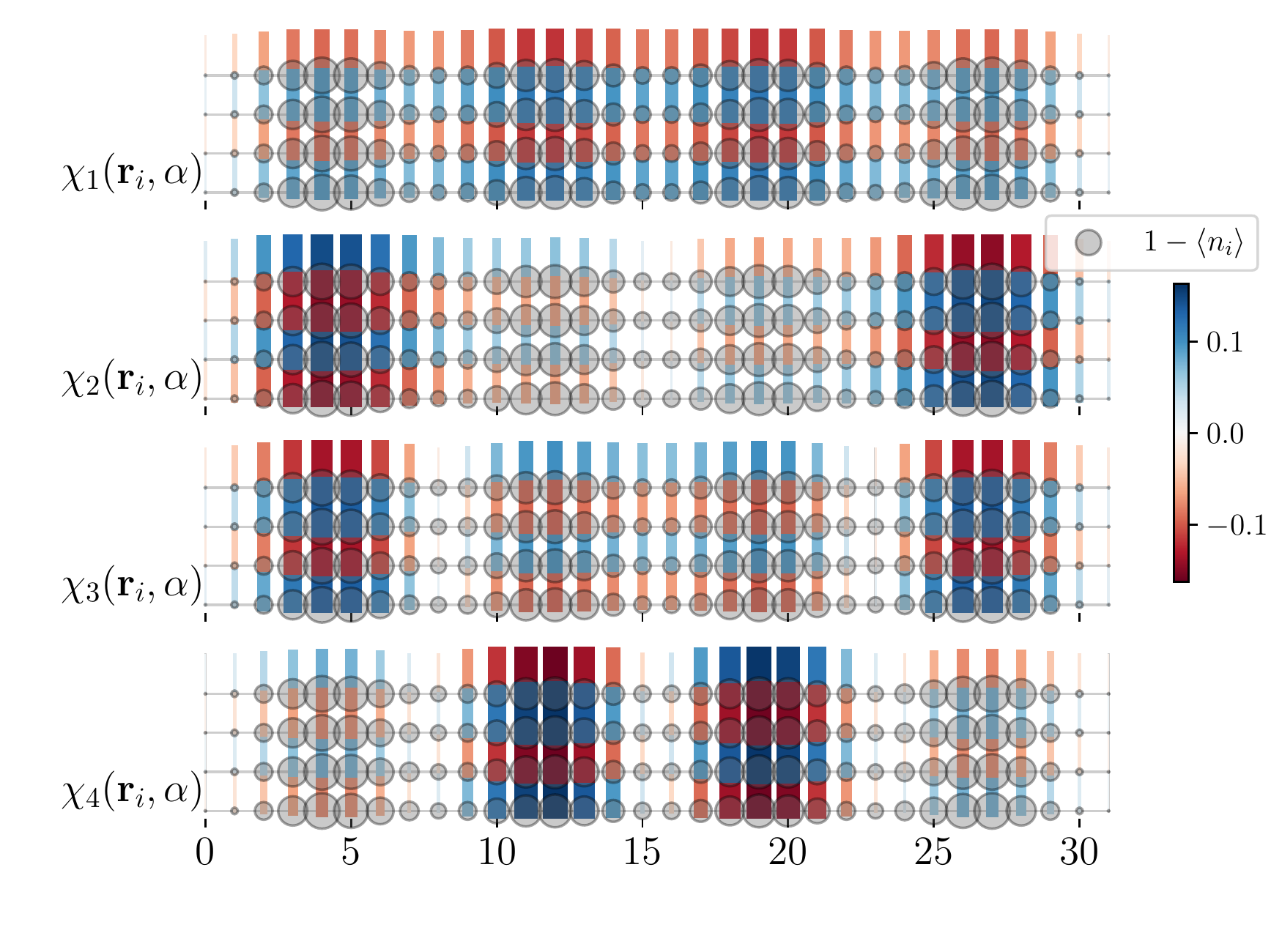}
    \caption{Condensate wave functions $\chi_n(\bm{r}_i, \alpha)$ for the dominant
    four eigenvalues of $\hat{\rho}_S$ on a $32\times 4$ cylinder at
    doping $p=1/16$ and $t^\prime= 0$. The hole-density $1 - \langle n_i \rangle$ 
    is shown as the area of the gray circles.}
    \label{fig:w4condwfsplaquette}
\end{figure}

The structure of the four dominant macroscopic wave functions 
$\chi_n(\bm{r}_i, \alpha)$ on the $W=4$ cylinder at $p=1/16$ and $t^\prime=0.2$ is
shown in in \cref{fig:w4condwfs}(a). The wave functions $\chi_n(\bm{r}_i, \alpha)$
depend both on the position $\bm{r}_i$ as well as the nearest-neighbor direction
$\alpha$. When $\alpha=\hat{\bm{x}}$ we show the value of $\chi_n(\bm{r}_i, \alpha)$ 
to the lattice edge right of site $\bm{r}_i$, if $\alpha=\hat{\bm{y}}$ it is shown
on the edge on top of site $\bm{r}_i$. We also show the local density of holes,
$1-\langle n_i \rangle$ superimposed. The most dominant condensate wave function
shown on top exhibits clearly extended uniform d-wave pattern, where horizontal and vertical bonds have opposite signs. The other two dominant modes exhibit a uniform d-wave pattern on a single stripe, while the
orientation and amplitude modulates between different stripes. A possible interpretation would be that uniform condensates form along the stripes of the system, which
hybridize by tunneling through a barrier of higher electron density. Hence, the
"fragments" of the condensate are individual condensates living on the stripes.
To demonstrate the relation between the condensates and the stripes more clearly,
we show the rung-averaged $d$-wave condensate wave function,
\begin{equation}
    \label{eq:rungavgdwave}
    \bar{\chi}_n^d(r_x) = \sum_{y=1}^W 
    \chi_n\left((r_x, r_y), \hat{\bm{x}}\right) - 
    \chi_n\left((r_x, r_y), \hat{\bm{y}}\right),
\end{equation}
in \cref{fig:w4condwfs}(b) alongside the rung-averaged hole density $1-\langle\bar{n}_i\rangle$, where $\bar{n}_i = \frac{1}{W}\sum_{j=1}^W n_{(x_i, y_j)}$.

We observe that the modulations of $\bar{\chi}_n^d(r_x)$
correspond exactly to the modulations in the charge density.

\begin{figure}
    \centering
    \includegraphics[width=\columnwidth]{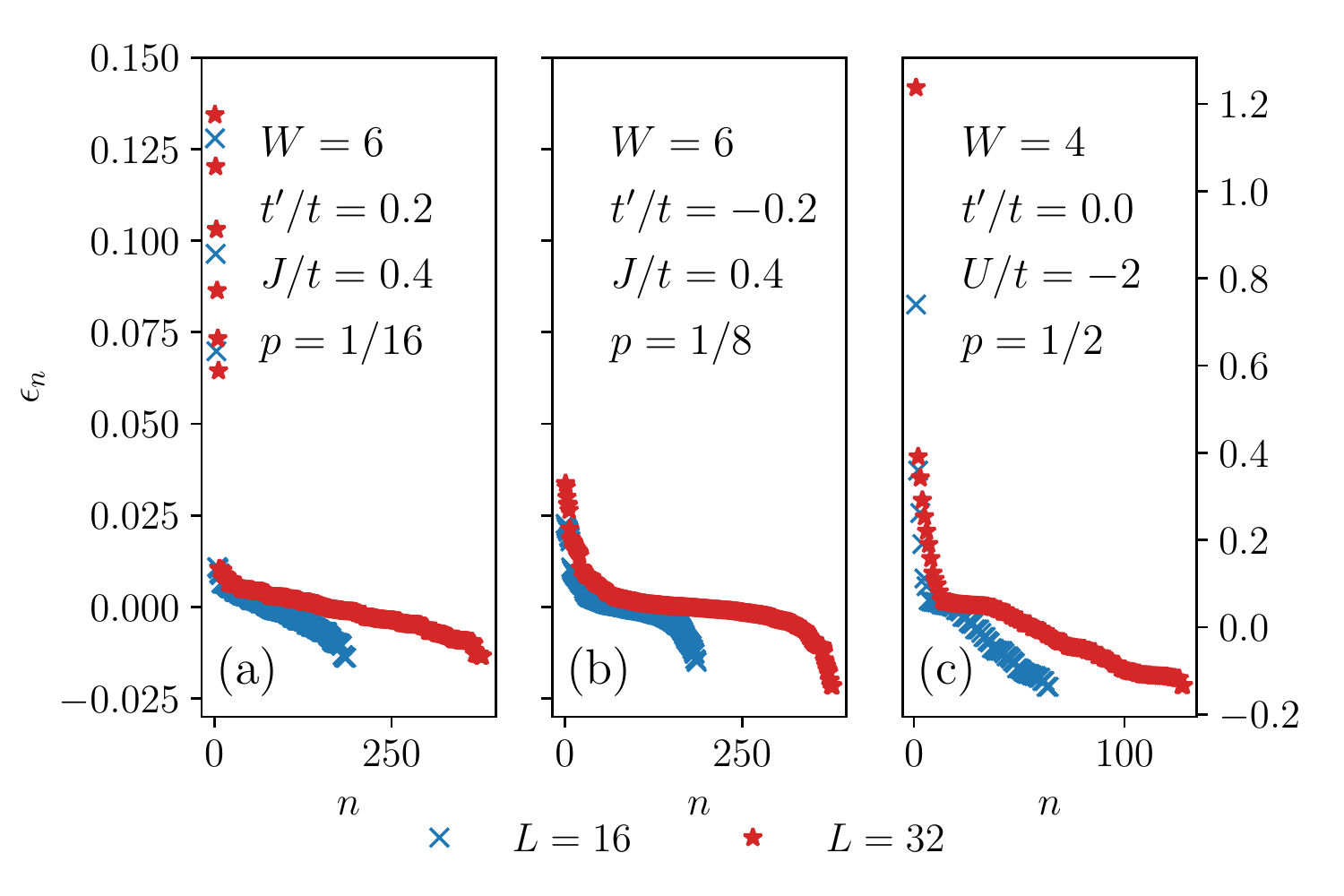}
    \caption{Spectrum $\varepsilon_n$ of $\hat{\rho}_S$ on the $W=6$ cylinder for (a) the d-wave superconducting state at $t^\prime=0.2$, $p=1/16$ and (b) a non-superconducting stripe state at $t^\prime=-0.2$, $p=1/8$. No dominant eigenvalues are observed in the non-superconducting case in (b). (c) Spectrum of $\rho_{\rm loc}$ (\cref{eq:rholoc}, diagonal elements have been set to zero) for the s-wave superconducting state realized in the attractive Hubbard model on a $W=4$ cylinder for $U/t=-2$, $p=1/2$, and $t^\prime/t=0$. Only one dominant eigenvalue is observed for this uniform condensate.} 
    \label{fig:pairing_cond_eigs_w6}
\end{figure}

Next, we show that the fragmentation of the condensate is not just a
particular feature of the LE2 phase on the $W=4$ cylinder but is more generic. 
We consider a different superconducting phase, which is stabilized on the $W=4$
cylinder, the plaquette-pairing phase at $t^\prime=0$ and $p=1/16$~\cite{Chung2020}, referred to as LE1 phase in Ref.~\cite{Jiang2020}.
The plaquette-pairing phase is a peculiarity of the width $W=4$ cylinder, where 
pairing is formed along the four-site plaquettes of the cylinder and is  different from the typical $d$-wave pairing state. The spectrum $\varepsilon_n$ of $\hat{\rho}_S$ closely resembles the case $t^\prime=0.2$, and the exact same number of dominant eigenvalues is observed. The condensate wave functions are shown in
\cref{fig:w4condwfsplaquette}(a). We clearly observe a plaquette pairing pattern, 
where the sign of $\chi_n(\bm{r}, \alpha)$ alternates in the $y$-direction,
while pairing along the $\hat{\bm{x}}$ direction is suppressed. Similar
to the $d$-wave condensates in \cref{fig:w4condwfs}, $\chi_n(\bm{r}, \alpha)$
is modulated by the stripes of the system.


\begin{figure}[t]
    \centering
    \includegraphics[width=\columnwidth]{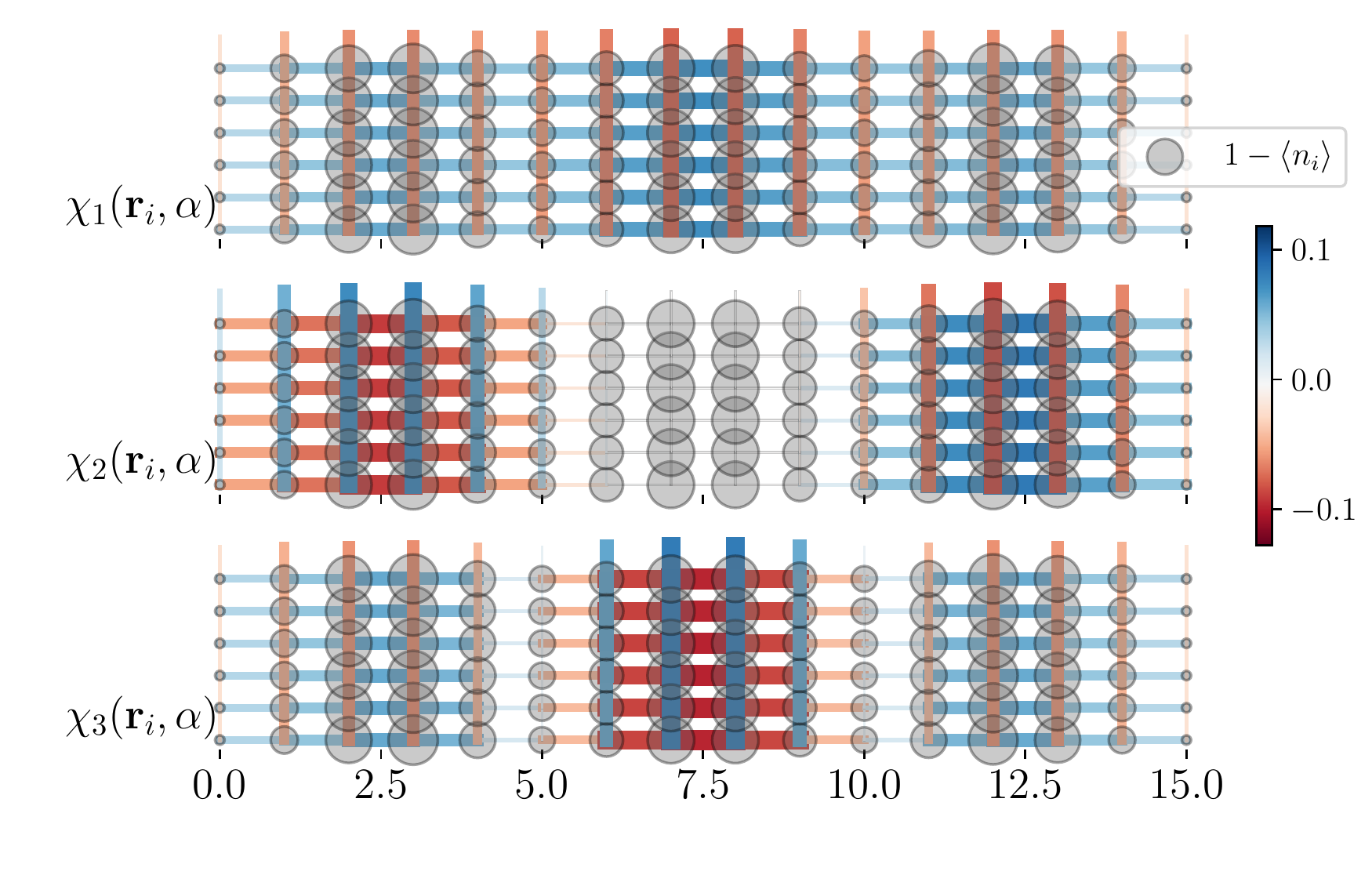}
    \caption{Condensate wave functions $\chi_n(\bm{r}_i, \alpha)$ for the dominant
    three eigenvalues of $\hat{\rho}_S$ on a $16\times 6$ cylinder at
    doping $p=1/16$ and $t^\prime= 0.2$. The hole-density $1 - \langle n_i \rangle$ 
    is shown as the area of the gray circles.}
    \label{fig:w6condwf}    
\end{figure}

The physics of the width $W=6$ cylinder is different from the $W=4$ cylinder in certain aspects. As established in Refs.~\cite{Jiang2021,HCJiang2021,Gong2021}, for $t^\prime < 0$ no superconductivity is observed and a charge density wave is stabilized. However, at small to intermediate doping and finite $t^\prime > 0$, a superconducting phase has been found. In \cref{fig:pairing_cond_eigs_w6} we show the spectrum of $\hat{\rho}_S$ in both the superconducting phase at $t^\prime=0.2$ and $p=1/16$ in panel (a) as well as the non-superconducting stripe phase at $t^\prime=-0.2$ and $p=1/8$ in panel (b). Only in the superconducting phase do we observe dominant eigenvalues, whose number again exactly matches the number of charge stripes. Therefore, the observation of dominant eigenvalues $\varepsilon_n$ is clearly associated with the superconductivity and not just the stripe order of the system. The associated macroscopic wave functions to the three dominant eigenvalues for $t^\prime=0.2$ and $p=1/16$ on the $16 \times 6$ cylinder are shown in \cref{fig:w6condwf}. Again, we observe a uniform d-wave pattern in the leading eigenvalue, which is modulated in the other two eigenvalues. To assess the stability of the fragmentation in the two-dimensional limit we compare the gap $\delta$ between the smallest dominant eigenvalue and the largest non-dominant eigenvalue between the $W=4$ and $W=6$ cylinders. We computed $\delta=0.041$ on the $32 \times 4$ cylinder and $\delta=0.054$ on the $32 \times 6$ cylinder. Hence, the gap is increasing with cylinder width, which is an indication of the stability of the condensate in the two-dimensional limit. 

We also consider the case of uniform s-wave superconductivity without the formation charge density wave. Such a state is realized in the attractive (negative-$U$) Hubbard model on the square lattice~\cite{Scalettar1989,Paiva2004}. Due to a difference in the pairing mechanism we consider the site-local pairing density matrix,
\begin{equation}
\label{eq:rholoc}
    \rho_{\rm loc}(\bm{r}_i | \bm{r}_j) = \langle \Delta_{\bm{r}_i}^\dagger \Delta_{\bm{r}_j}\rangle \quad \rm{ where } \quad  \Delta_{\bm{r}_i}^\dagger = c^\dagger_{\bm{r}_i\uparrow} c^\dagger_{\bm{r}_i\downarrow}.
\end{equation}
\cref{fig:pairing_cond_eigs_w6}(c) shows that a single dominant eigenvalue is formed at $U/t = -2$ and $t^\prime/t=0$ on the W=4 cylinder at quarter filling, i.e. $p=1/2$ increasing with system size.  

\begin{figure}
    \centering
    \includegraphics[width=\columnwidth]{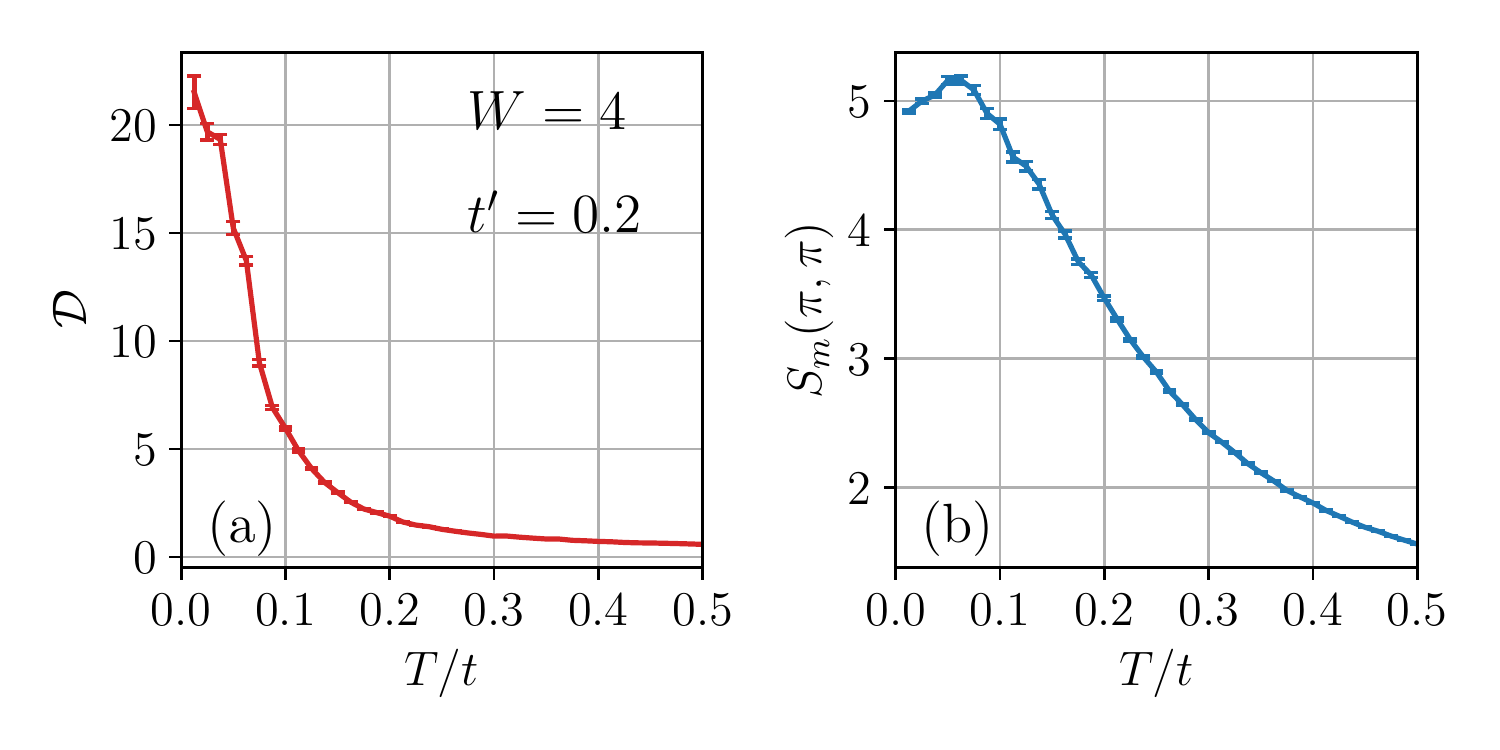}
    \caption{Temperature dependence of the d-wave pairing susceptibility $\mathcal{D}$ (a) and the antiferromagnetic spin structure factor $S_m(\pi,\pi)$ (b) for $t-t^\prime -J$ model on the $32 \times 4$ cylinder at $p=1/16$ and $t^\prime=0.2$. Strong pairing correlations develop below a temperature $T/t \approx 0.05$, which suppresses antiferromagnetism.}
    \label{fig:thermo}
\end{figure}
Finally, we investigate how the fragmented condensate can emerge from a "normal" state. Therefore, we study the temperature dependence of the d-wave pairing susceptibility,
\begin{equation}
\mathcal{D} = \sum_{\alpha, \beta}(-1)^{\alpha\cdot\beta + 1}\sum_{ \mathbf{r}_i, \mathbf{r}_j} \hat{\rho}_S(\mathbf{r}_i, \alpha | \mathbf{r}_j, \beta),
\end{equation}
and the magnetic structure factor $S_m(\mathbf{q})$ at the antiferromagnetic ordering vector $\mathbf{q}=(\pi,\pi)$ using the METTS method with maximal bond dimension $D=2000$~\cite{Wietek2021,Wietek2021b}. We observe that strong pairing correlations develop below a temperature of $T/t\approx 0.05$. Antiferromagnetic correlations develop at a higher temperature but are finally suppressed by pairing correlations. 

\section{Discussion and conclusion}
Our results suggest a simple physical picture of the interplay of stripe order and superconductivity. Individual superconducting condensates are formed on the stripes of the system and hole-pairs can tunnel through a barrier given by the maxima in the electron density. The superconducting stripes could thus be regarded as an emergent array of Josephson junctions. While we found the most dominant macroscopic wave function to be a uniform superposition of the condensate fragments, it is an important open question under which circumstances different modes, e.g. a $\pi$-phase shift Josephson junction, could be realized as the dominant contribution. Moreover, the smallest dominant eigenvectors shown in \cref{fig:w4condwfs} ($n=4$) and \cref{fig:w6condwf} ($n=3$) are pair-density waves~\cite{Agterberg2020}, where the condensate wave function is modulated from stripe to stripe. Such states have previously been suggested for the $t$-$t^\prime$-$J$ model from variational Monte Carlo simulations~\cite{Himeda2002}. Interestingly, recent experiments on \ch{La_{2–x}Ba_{x}CuO4} have highlighted the possibility of having pair correlations within stripes without coherence between the stripes~\cite{Yangmu2019,Wardh2022}. This observation could indeed be explained by the fragmentation of the superconducting state by stripes, a fundamental mechanism we have now revealed in the $t$-$t^\prime$-$J$ model. 


\section*{Acknowledgements}
I am very grateful for insightful discussions with Andrew Millis, Steven R. White and Antoine Georges. The DMRG results obtained using the ITensor Library ~\cite{itensor}. The Flatiron Institute is a division of the Simons Foundation.

\bibliography{main.bib}

\end{document}